\documentclass[letter,twocolumn]{jpsj3}
\usepackage{txfonts}

\title{Monte Carlo Study of Magnetoresistance in a Chiral Soliton Lattice}

\author{Shun Okumura\thanks{s.okumura@aion.t.u-tokyo.ac.jp}, Yasuyuki Kato, and Yukitoshi Motome}
\inst{Department of Applied Physics, The University of Tokyo, Hongo, Bunkyo, Tokyo 113-8656, Japan} 

\abst{
  Monoaxial chiral magnets can form a peculiar noncollinear spin structure called the chiral soliton lattice in an applied magnetic field perpendicular to the helical axis. 
We study magnetic properties and electrical transport in the chiral soliton lattice using a Monte Carlo simulation for a one-dimensional Kondo lattice model including the Dzyaloshinskii--Moriya interaction between classical localized spins.
We show that the model exhibits a helical spin structure at a zero magnetic field, which turns into the chiral soliton lattice, and finally, to a forced ferromagnetic state with an increase of the external magnetic field. In the chiral soliton lattice state, we find negative magnetoresistance proportional to the number of solitons at low temperature, which corroborates the spin scattering of electrons by chiral solitons. We also discuss the temperature and magnetic field dependence of the spin structure factor and electrical resistivity, in comparison with experiments for CrNb$_{3}$S$_{6}$.
}

\begin{document}
\maketitle

Chirality plays an important role in a broad area of materials science. It is characterized by reflection without rotational symmetry, often termed as handedness. For instance, chirality has long been studied for molecular structures in chemistry and biology, where chemical activities strongly depend on the handedness. Chiral structures in solids have also attracted considerable attention in condensed matter physics. The archetypal examples are found in chiral magnets. Chiral magnetic textures such as spin spirals play a decisive role in the electronic properties of solids, such as optical responses, magnetoelectric behaviors, and transport properties. For example, a noncoplanar swirling magnetic structure called skyrmion\cite{Nagaosa2013,Rossler2006,Muhlbauer2009,Yu2010,Heinze2011} leads to an unconventional anomalous Hall response originating from the spin Berry phase mechanism related to the topological nature of the spin texture\cite{Lee2009,Nueubauer2009,Schulz2012}.

Recently, interesting magnetotransport phenomena have been observed in a magnetic conductor with a chiral magnetic structure, CrNb$_3$S$_6$. This compound exhibits a monoaxial helical spin order at low temperature $(T)$ in the absence of a magnetic field\cite{Moriya1982,Miyadai1983}. When the magnetic field is applied perpendicular to the helical axis, the helical order turns into a chiral soliton lattice (CSL)\cite{Togawa2012}. The CSL is a spin texture similar to a periodically twisted ribbon, whose period increases as the magnetic field increases. In the CSL phase, negative magnetoresistance is observed with peculiar field dependence, which appears to correlate with the period of CSL\cite{Togawa2013}. On the other hand, electrical resistivity increases with $T$\cite{Togawa2013}, presumably due to the thermal disturbance of the spin textures. Furthermore, a domain structure with opposite crystalline chirality leads to a discrete change of electrical resistivity.\cite{Togawa2015} All these experimental results indicate strong interplay between the peculiar magnetism and electrical transport in CrNb$_3$S$_6$\cite{Togawa2016}.

A pioneering work on CSL was carried out by Dzyaloshinskii in his seminal papers in 1960s\cite{Dzyaloshinskii1964,Dzyaloshinskii1965}; he developed the Ginzburg--Landau theory to discuss the critical properties before entering the forced ferromagnetic (FFM) state by increasing the magnetic field at $T=0$\cite{Dzyaloshinskii1964,Dzyaloshinskii1965}. Moriya and Miyadai introduced Dzyaloshinskii--Moriya (DM) interaction\cite{Dzyaloshinskii1958,Moriya1960} to a ferromagnetic Heisenberg model to explain a CSL in CrNb$_3$S$_6$\cite{Moriya1982}. Recently, the chiral magnetic properties of CrNb$_3$S$_6$ have been revisited by Kishine {\it et al.}\cite{Kishine2005,Kishine2015} on the basis of a one-dimensional continuum model, called the chiral sine--Gordon model; electrical transport in the CSL\cite{Kishine2011} and sliding dynamics of chiral solitons by an electrical current \cite{Bostrem2008-1,Bostrem2008-2,Borisov2009,Kishine2012} were discussed in the same framework. However, these studies were limited to $T = 0$ or in the vicinity of the phase boundary between the CSL and paramagnetic phases at finite $T$, because of the continuous approximation. For the finite-$T$ properties, three-dimensional localized spin models were studied by numerical simulations as well as mean-field approximations\cite{Shinozaki2016,Nishikawa2016,Laliena2016}, where the degree of freedom for itinerant electrons is not explicitly taken into account. Hence, the comprehensive understanding of the characteristic magnetoresistive behaviors in chiral magnets such as CrNb$_3$S$_6$ awaits theoretical studies of the systems with itinerant electrons. 

In this letter, in order to clarify the relation between the peculiar magnetic structure and electrical transport in the CSL state, we investigate the simplest itinerant electron model, i.e., a one-dimensional Kondo lattice model with the DM interaction, using Monte Carlo (MC) simulation where the localized spins are treated as classical spins. We show that the model exhibits a helical spin structure at a zero magnetic field, which is turned into a CSL by applying the field perpendicular to the chiral axis. 
By calculating the spin structure factor, optical conductivity, and winding number corresponding to the number of chiral solitons, we find that the coherent weight of the optical conductivity, which is a measure of the electrical conductivity, increases along with the decrease of the chiral solitons. Moreover, while raising $T$, the CSL melts because of thermal fluctuations and the coherent conduction is suppressed. 
These results are qualitatively consistent with the experimental data for CrNb$_3$S$_6$\cite{Togawa2016}. 
To the best of our knowledge, this is the first theoretical result that explicitly indicates topological magnetic defects in the CSL affect the electrical transport via spin scattering. 

For capturing the essential physics of CSL in CrNb$_3$S$_6$, we consider an extension of the ferromagnetic Kondo lattice model that includes the degree of freedom of both itinerant electrons and localized magnetic moments. CrNb$_3$S$_6$ has a layered structure of NbS$_2$ intercalated with Cr. The $d$-electron levels at each Cr are split by crystalline electric fields into the higher-energy $e_g$ and lower-energy $t_{2g}$ levels. The localized moments in the model are attributed to rather localized $t_{2g}$ electrons. In CrNb$_3$S$_6$, the localized moments are strongly coupled with each other in a ferromagnetic manner within each layer\cite{Miyadai1983}, and hence, we can approximate them by a one-dimensional array of large magnetic moments treated as classical spins. Meanwhile, although the conduction bands might have complicated structures owing to the hybridization between the Cr $e_g$ orbitals and the NbS$_2$ $s$ and $p$ orbitals, we represent them by a single band for simplicity. We assume that the itinerant electrons are coupled with the classical spins by a ferromagnetic exchange coupling, which mimics the Hund's-rule coupling for $e_g$ electrons or the $s$-$d$ type coupling for $s$ and $p$ electrons. In addition, considering the chiral lattice structure, we introduce the DM interaction between the neighboring localized spins. This interaction originally comes from the relativistic spin-orbit coupling; however, here, we include it phenomenologically while omitting the spin-orbit coupling in the itinerant electrons. We note that the same-type DM interaction introduced to the localized spin models well explains the helical order and CSL\cite{Kishine2015,Shinozaki2016,Nishikawa2016,Laliena2016}. Then, the Hamiltonian for the extended Kondo lattice model is given by
\begin{align}
	H = &-t\sum_{l,\mu}(c^{\dagger}_{l\mu}c^{\;}_{l+1\mu}+\mathrm{h.c.})
        -J\sum_{l,\mu,\nu}c^{\dagger}_{l\mu}{\boldsymbol \sigma}_{\mu\nu}c^{\;}_{l\nu}\cdot{\mathbf S}_{l}\nonumber\\
        &-{\mathbf D}\cdot\sum_{l}{\mathbf S}_{l}\times{\mathbf S}_{l+1}-h\sum_{l}S_l^{x},
\label{eq:H}
\end{align}
where $c_{l\mu}(c^{\dagger}_{l\mu})$ is an annihilation (creation) operator for a $\mu$-spin electron at site $l$ on the one-dimensional chain ($\mu = \uparrow$ or $\downarrow$), and ${\mathbf S}_{l}=(S_l^x,S_l^y,S_l^z)$ is a three-component vector with normalized length $|{\mathbf S}_{l}|=1$ representing the large spin in each layer. The first term describes the kinetic energy of itinerant electrons; $t$ is a transfer integral between the nearest-neighbor sites. The second term is for the onsite coupling between the itinerant electrons and localized moments; $J$ is a positive coupling constant and $\boldsymbol{\sigma} = (\sigma^x,\sigma^y,\sigma^z)$ are the Pauli matrices. The third term represents the DM interaction with the DM vector ${\mathbf D}=D\hat{z}$, where $\hat{z}$ is a unit vector along the chain direction. The last term is the Zeeman coupling to the magnetic field $h$ perpendicular to the chain direction, taken as $\hat{x}$, which is introduced only to the localized moments for simplicity. 

We investigate the thermodynamic and transport properties of the model in Eq.~(\ref{eq:H}) by utilizing a MC method. 
In this method, we update the configuration of classical spins $\{\mathbf{S}_l\}$ using the Metropolis algorithm. 
The statistical weight is given by the free energy of itinerant electrons for a given configuration of $\{\mathbf{S}_l\}$, which is calculated by the exact diagonalization of the bilinear Hamiltonian in terms of $c_{l\mu}^\dagger$ and $c_{l\mu}$. In the following, we show the results for the system with $N=110$ sites under the open boundary condition; we confirm that the finite-size effects are sufficiently small for our discussions. 
We compute the thermodynamic properties using the grand canonical ensemble with the electron filling at quarter filling (half electron per site on average), while the results do not change qualitatively for other generic fillings. 
We perform $4.8\times 10^5$ measurements after $10^4$ MC steps for thermalization. We divide the MC samples into 48 bins to estimate the statistical errors by the standard deviation among the bins. 

In the following calculations, we set the energy unit as $t = 1$ and take $J = 2$. We note that this leads to a ferromagnetic ground state in the absence of the DM interaction at quarter filling. We choose $D = 0.035$, which stabilizes a helical state with the period of 10 sites at $h=0$, as shown below. The following results remain the same qualitatively, when we take a smaller $D$ to reproduce the longer period in CrNb$_3$S$_6$; however, a much larger system size is necessary for the numerical simulation. In the current model in Eq.~(\ref{eq:H}), there is a relation between the helical pitch and the model parameters $J$, $D$, and electron filling, which can be obtained by a variational argument for the ground state; the details will be reported in a future study.

\begin{figure}[b]
\centering
\includegraphics[width=\columnwidth,clip]{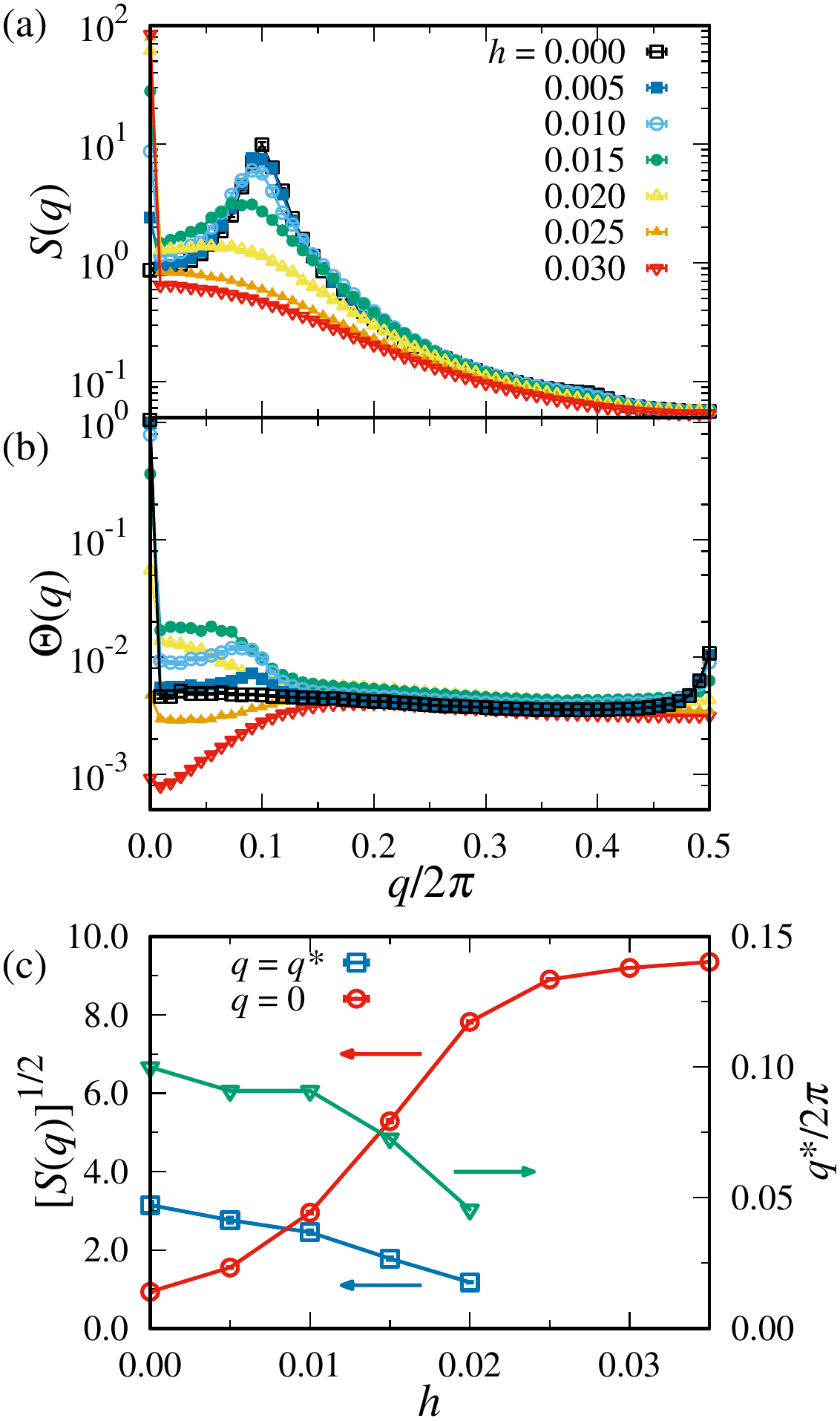}
\caption{(a) Spin structure factor $S(q)$ and (b) structure factor for the relative angles of neighboring spins $\Theta(q)$ obtained by MC simulation for the model in Eq.~(\ref{eq:H}) while varying the external magnetic field $h$. (c) Square root of $S(q^{*})$ and $S(0)$ as functions of $h$. $q^{*}$ is the nonzero wave number at which $S(q)$ shows a peak. The data are calculated at $T = 0.01$.
}
\label{f1}
\end{figure}

First, we show the MC results in the absence of an external magnetic field, $h=0$. 
Our model in Eq.~(\ref{eq:H}) exhibits a helical spin structure at low $T$ (the long-range order is not allowed at finite $T$ in the present one-dimensional model). 
Figure~\ref{f1}(a) presents the spin structure factor for the localized spins at low $T=0.01$, which is defined by
\begin{align}
	S(q) = \frac{1}{N}\sum_{l,m}e^{i(l-m)q}\langle{\mathbf S}_{l}\cdot{\mathbf S}_{m}\rangle.
\end{align}
$S(q)$ shows a single peak at $q = 2\pi/10$. We also compute the structure factor for the relative angles between neighboring spins projected on the $xy$ planes, defined as
\begin{align}
	\Theta(q) = \frac{1}{N-1}\sum_{l,m}e^{i(l-m)q}\langle\Delta\theta_{l}\Delta\theta_{m}\rangle,
	\label{eq:Theta_q}
\end{align}
where $\Delta\theta_{l}$ is the relative angle between $l$th and $(l+1)$th localized spins: 
\begin{align}
\Delta\theta_l = \sin^{-1} \frac{S^{x}_l S^{y}_{l+1} - S^{y}_l S^{x}_{l+1}}{\sqrt{(1-S^{z2}_{l})(1-S^{z2}_{l+1})}}.
\label{eq:Delta_theta}
\end{align}
At $h=0$, $\Theta(q)$ shows a peak only at $q=0$ as shown in Fig.~\ref{f1}(b), indicating that the relative angles are uniform along the chain.\cite{Footnote} 
The results of $S(q)$ and $\Theta(q)$ show that the system exhibits a helical spin structure at $h=0$ with the period of 10 sites.

Next, we show the results in an applied magnetic field, $h > 0$. As shown in Fig.~\ref{f1}(a), the peak in $S(q)$ shifts to a smaller $q$ with a slightly decreasing peak height (we denote the wave number as $q^*$), and simultaneously, another peak grows at $q=0$ as $h$ increases. The behaviors are summarized in Fig.~\ref{f1}(c).  
On the other hand, as shown in Fig.~\ref{f1}(b), the $q=0$ peak in $\Theta(q)$ is reduced, whereas another peak is developed at $q^{*}$. These results consistently indicate that the helical spin structure at $h=0$ changes into a CSL by introducing $h$. The period of the CSL increases as $h$ increases, which is given by $2\pi/q^*$ [see Fig.~\ref{f1}(c)]. 

With a further increase in $h$, we find that the CSL turns into a FFM state around $h = 0.025-0.030$, where $q^*$ approaches zero and $S(0)$ saturates. We note that the crossover occurs at $h$ comparable to the strength of the DM interaction $D = 0.035$.

\begin{figure}[t]
\centering
\includegraphics[width=\columnwidth,clip]{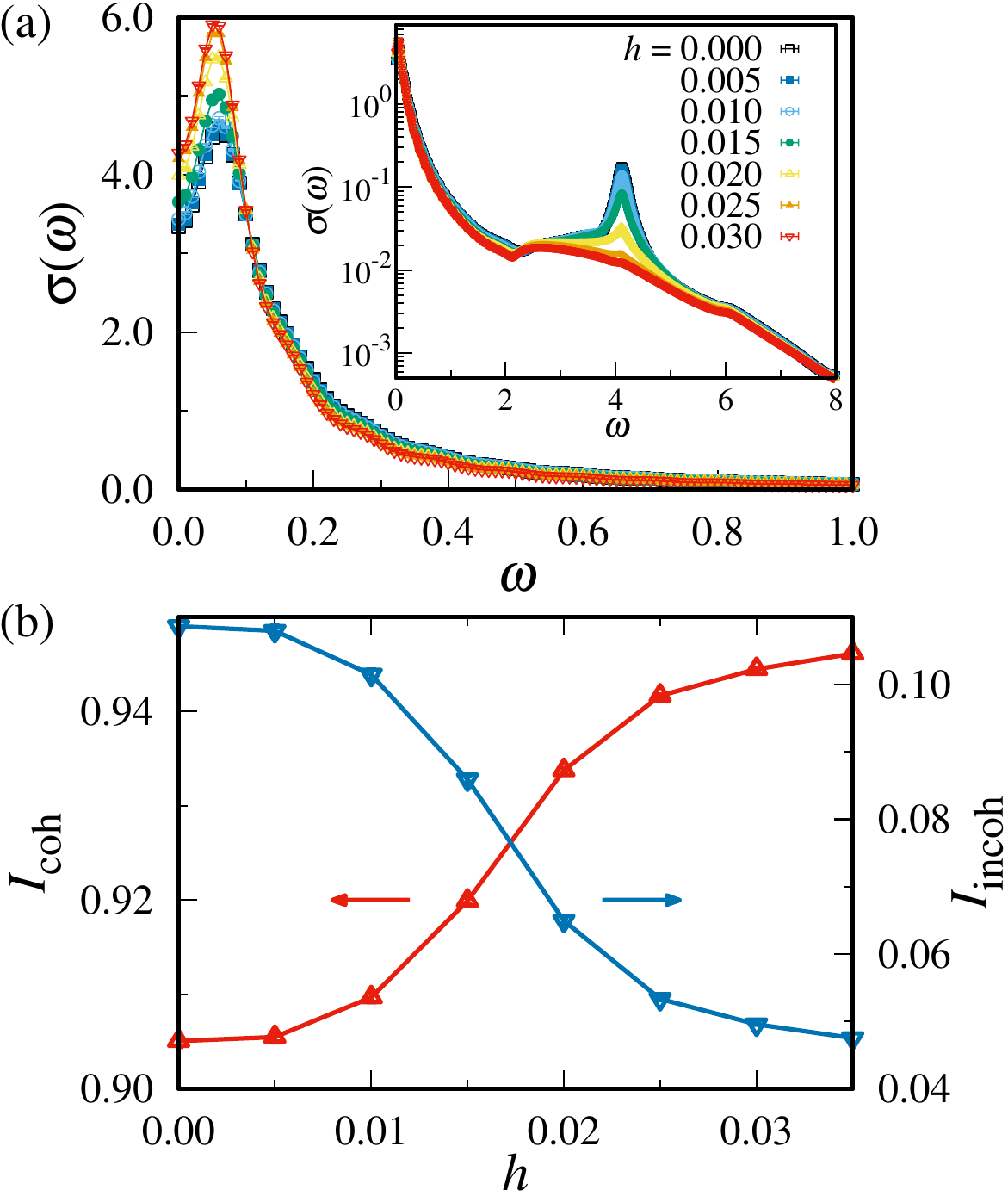}
\caption{(a) Optical conductivity $\sigma(\omega)$ in the low-$\omega$ part at $T = 0.01$ for several $h$. The inset presents the overall behavior in a wider range of $\omega$. (b) Integrated coherent and incoherent weights, $I_{\rm coh}$ and $I_{\rm incoh}$, respectively, of the optical conductivity as functions of $h$.}
\label{f2}
\end{figure}

\begin{figure}[b]
\centering
\includegraphics[width=\columnwidth,clip]{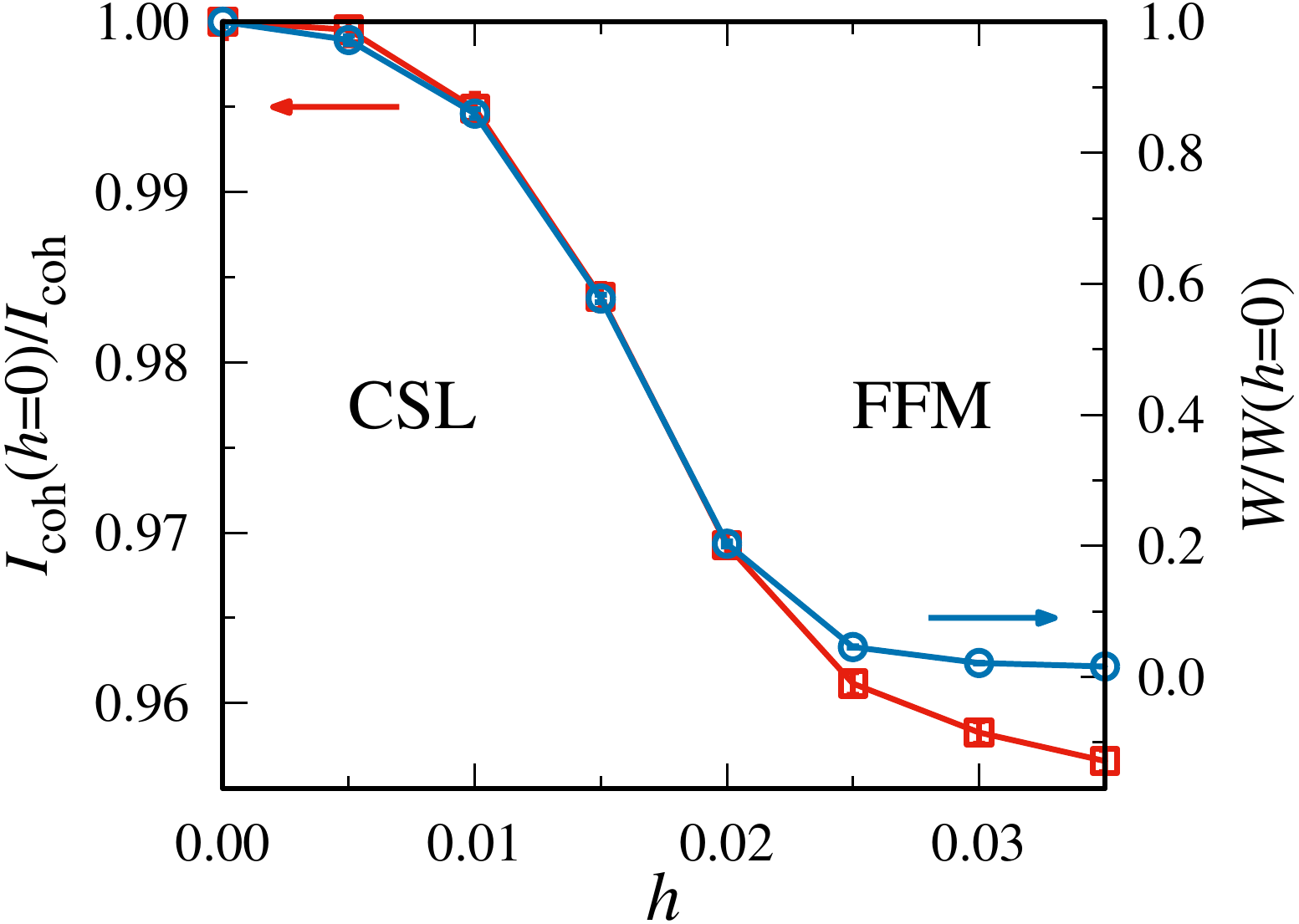}
\caption{The inverse of the coherent weight of the optical conductivity, $I_{\rm coh}^{-1}$, and the winding number $W$ of the spin structure as functions of $h$. Both are normalized by the values at $h = 0$. The data are calculated at $T = 0.01$.
}
\label{f3}
\end{figure}

Now, we investigate the electrical transport in an applied magnetic field. For this purpose, we compute the optical conductivity by using the Kubo formula
\begin{align}
	\sigma(\omega) = \Bigg\langle\mathrm{Re}\left[\frac{i}{N}\sum_{k,l}\frac{f(\varepsilon_{k})-f(\varepsilon_{l})}{\varepsilon_{l}-\varepsilon_{k}}\frac{|\langle l|j_{z}|k\rangle|^{2}}{\omega -(\varepsilon_{k}-\varepsilon_{l})+i\delta}\right]\Bigg\rangle,
\end{align}
where $\varepsilon_{k}$ and $|k\rangle$ are the $k$th eigenvalue and eigenvector, respectively, of the bilinear Hamiltonian in Eq.~(\ref{eq:H}) for a spin configuration $\{\mathbf{S}_l\}$ generated by MC sampling; $f(\varepsilon_{k})$ is the Fermi distribution function, and $j_{z}$ is the current operator in the {\it z} direction given by $j_{z} = -it\sum_{l,\mu}(c^{\dagger}_{l\mu}c^{\;}_{l+1\mu}-c^{\dagger}_{l+1\mu}c^{\;}_{l\mu})$\cite{Yunoki1998}. We take $\delta = 1/22$. 
The MC result is plotted in Fig.~\ref{f2}. $\sigma(\omega)$ shows two peaks: one is around $\omega=0$ and the other is at $\omega \simeq 2J = 4$. 
The former originates from the coherent motion of electrons, although the peak is slightly shifted from $\omega=0$ because of the finite-size effect (the peak approaches $\omega=0$ as $N\to \infty$). Meanwhile, the latter peak represents an incoherent component originating from the interband matrix elements of $j_z$ since the bands are split by $\sim2J$.

To provide a measure of the conductivity of the system, we estimate the coherent weight of $\sigma(\omega)$ by the integral of the low-$\omega$ part as
\begin{align}
	I_{\rm coh} = \int_{0}^{\omega_{0}}\sigma(\omega)d\omega.
\end{align}
In the present calculations, we set the cutoff $\omega_{0} = J = 2$.
The result is shown in Fig.~\ref{f2}(b). 
The coherent weight grows with increasing $h$, corresponding to the development of the ferromagnetic component $S(0)$ in the CSL plotted in Fig.~\ref{f1}(c). 
This indicates that the system exhibits negative magnetoresistance.
We also calculate the incoherent weight by the integral of the high-$\omega$ part as
\begin{align}
I_{\rm incoh} = \int_{\omega_{0}}^{\omega_{1}} \sigma(\omega) d\omega, 
\end{align}
where we set $\omega_{1} = 4J = 8$ to include the contribution from the incoherent peak at $\omega\simeq 2J$. 
In stark contrast to $I_{\rm coh}$, $I_{\rm incoh}$ decreases while $h$ increases, as shown in Fig.~\ref{f2}(b). These results clearly indicate a close relation between the systematic change in the magnetic state and the electrical transport: while $h$ increases, the period of CSL grows longer, the ferromagnetic component becomes more dominant, and the coherent electron motion is enhanced.
This suggests that the spin scattering of itinerant electrons by the peculiar spin defects, i.e., chiral solitons, plays a key role in the electrical transport. 
 
 In order to confirm the spin scattering picture, we compute the number of chiral solitons and compare it with electrical resistivity. The number of chiral solitons is defined by the winding number of the one-dimensional spin configuration as
\begin{align}
	W = \frac{1}{2\pi}\sum_{l}\langle\Delta\theta_{l}\rangle,
\end{align}
where $\Delta\theta_{l}$ is defined in Eq.~(\ref{eq:Delta_theta}). As shown in Fig.~\ref{f3}, the MC result for the winding number $W$ correlates well with the inverse of the coherent weight of $\sigma(\omega)$, $I_{\rm coh}^{-1}$. 
As $I_{\rm coh}$ gives a measure of the conductivity of the system, the result indicates the electrical resistivity is proportional to the number of chiral solitons in the CSL state. This confirms that the electrical transport is governed by the scattering of electrons by the chiral solitons in the present system. 
A similar statement was inferred in the previous studies~\cite{Kishine2015,Togawa2016}; however, our calculations demonstrate the relation by a direct calculation of the electrical transport for the first time, to the best of our knowledge.

\begin{figure}[b]
\centering
\includegraphics[width=\columnwidth,clip]{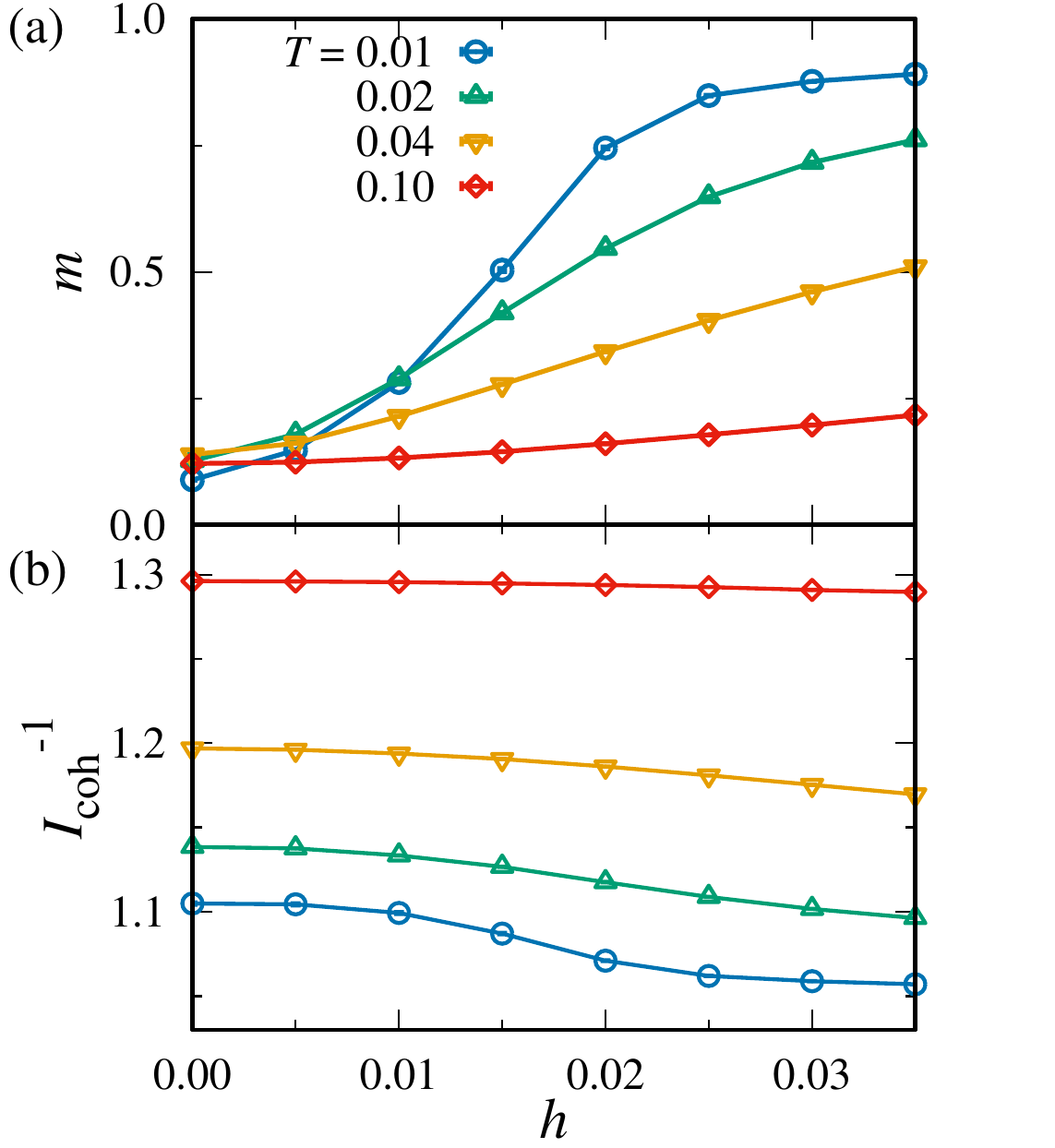}
\caption{(a) Magnetization $m$ and (b) the inverse of the coherent conductivity $I_{\rm coh}^{-1}$ as functions of $h$ at several temperatures.}
\label{f4}
\end{figure}

Finally, we investigate the $T$ dependence of the spin structure and the measure of the electrical resistivity, $I_{\rm coh}^{-1}$. Figure~\ref{f4}(a) 
illustrates the ferromagnetic moment $m=\sqrt{S(0)/N}$. 
At low $T$, $m$ increases rapidly while $h$ increases, as the helical state at $h=0$ is turned into the FFM state through the CSL. As $T$ increases, the growth of $m$ is reduced by thermal fluctuations. Correspondingly, $I_{\rm coh}^{-1}$ is enhanced by raising $T$, as shown in Fig.~\ref{f4}(b). 
At the same time, the $h$ dependence becomes weaker, indicating that negative magnetoresistance becomes weaker for higher $T$. These MC results qualitatively explain the experimental results for CrNb$_{3}$S$_{6}$\cite{Togawa2013}. 

In summary, we investigated the magnetic properties and electrical transport phenomena in the CSL using the MC simulation for the one-dimensional Kondo lattice model with the DM interaction between neighboring localized spins.
By appropriately considering the degree of freedom of not only localized spins but also itinerant electrons, to the best of our knowledge, this study is the first to demonstrate that electrical resistivity is proportional to the number of chiral solitons at low $T$.
We also clarified that, while $T$ increases, the resistivity is enhanced with a decrease of negative magnetoresistance, corresponding to the disturbance of the spin texture of CSL by thermal fluctuations. 
The results clearly demonstrate that the spin scattering of itinerant electrons by chiral solitons is essential for the electrical transport in the present system.
Our results qualitatively reproduce the experimental results in a chiral magnetic conductor, CrNb$_3$S$_6$, except for the critical behaviors associated with the finite-$T$ phase transitions.
To discuss the finite-$T$ phase transitions by varying $T$ and $h$, it is necessary to extend the model to three dimensions. 
Such an extension requires much larger computational cost; nevertheless, it is feasible to achieve this by utilizing sophisticated numerical techniques such as the Langevin dynamics with the kernel polynomial method\cite{Barros2013,Ozawa2016,Ozawa2017}. 
This is left for future work.

\begin{acknowledgment}
S.~O. thanks to R.~Ozawa for his advice on the MC algorithm. 
The authors thank J.~Kishine, Y.~Nishikawa, R.~Ozawa, M.~Shinozaki, and Y.~Togawa for fruitful discussions.
This research was supported by KAKENHI (No.~15K05176).
Part of the computation in this work was carried out at the Supercomputer Center, Institute for Solid State Physics, the University of Tokyo.

\end{acknowledgment}


\begin{thebibliography}{9}
\bibitem{Footnote} The small peak at $q=\pi$ originates from the finite-size effect, which is reduced while increasing $N$.
\end{thebibliography}
\end{document}